\documentclass[sigconf]{acmart}
\usepackage{booktabs,tabularx,array, xspace,subcaption}
\newcolumntype{L}[1]{>{\raggedright\arraybackslash\hsize=#1\hsize}X}

\setcopyright{none}
\copyrightyear{2026}
\acmYear{2026}
\acmDOI{}
\acmConference[LOCO '26]{Proceedings of the 2nd International Workshop on Low Carbon Computing (LOCO 2026)}{Sep 10--11,
  2026}{Lancaster \& Online}
\acmISBN{}

\newcommand{\ceq}{CO$_2$e\xspace}

\begin{document}

\title[Hot Games]{Hot Games:  Towards a Holistic Assessment of the Planet Warming Emissions of Video Games based on 2024--2025 Data}

\author{Mike Hazas}
\email{mike.hazas@it.uu.se}
\orcid{0000-0002-9732-0283}
\author{Kevin Charles Dalli}
\email{kevin.dalli@it.uu.se}
\author{Arjun Menon}
\orcid{0000-0002-6711-0584}
\email{arjun.menon@it.uu.se}
\author{Ossian Nordgren}
\email{ossian.nordgren@it.uu.se}
\affiliation{%
  \institution{Department of Information Technology, \\ Uppsala University}
  \country{Sweden}
}
\author{Ben Abraham}
\email{ben@sustainablegamesalliance.org}
\orcid{}
\affiliation{%
  \institution{Sustainable Games Alliance}
  \city{Melbourne}
  \country{Australia}
}

\begin{abstract}
Following on recent reports on specific platforms or companies, this paper provides an assessment of the global impact of the production and use of video games.  It draws together publicly available data on game development, hardware, games sold, download sizes, time spent playing games on different platforms, and subscriptions to multiplayer and cloud game services.  It provides an update to figures published 2020 and 2022.  Crucially, our account of emissions related to video games considers a wide range of categories, yet contains enough detail to be critiqued and improved in the future.
\end{abstract}


\settopmatter{printacmref=false}
\maketitle

\section{Introduction}

The energy and environmental impact of making and playing video games has received increasing attention lately, as evidenced by organisations such as the Sustainable Games Alliance and Playing for the Planet; and industry efforts such as the Xbox Sustainability Toolkit.  Despite rising expectations for industry reporting~\cite{fors2025-reporting}, the existing accounts tend to be focused on one category or company, and are outdated.

Assessing global carbon equivalent emissions of video games enables the industry to be better compared to other digital home entertainment (e.g., video streaming) and sectors commonly associated with climate impacts (transportation, space heating). Broad, yet detailed estimations also provide a wider overview for identifying especially impactful components of the video game life cycle, insights which can inform future policy and industry practices to lessen overall emissions. Finally, our account helps to contextualise claims made by the video games industry of their sustainability and low climate impact~\cite{Dataspelsbranschen2024}.

\begin{figure}
    \centering
    \includegraphics[width=\linewidth]{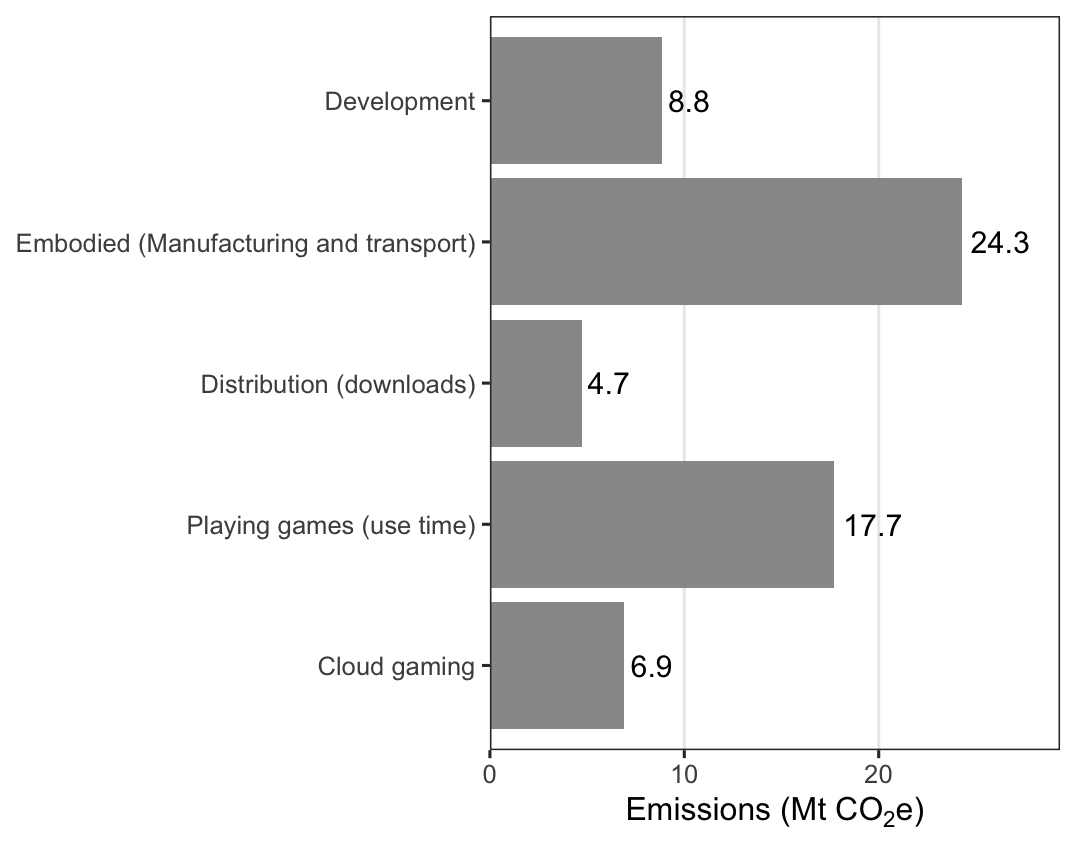}
    \caption{Annual emissions by category. Total 62.5 Mt \ceq}
    \label{fig:emissions_by_source}
\end{figure}

This paper provides a revision to some of the estimates appearing in a 2020 paper by Marsden et al.~\cite{marsden2020}; and serves as corollary to Abraham's 2022 book~\cite{abraham2022-digitalgamesafterclimatechange}.  The present paper takes a broad consideration of the recent devices and usage associated with video games.  Drawing from publicly available sources, we estimated (fig.~\ref{fig:emissions_by_source}) emissions arising from the development of games, manufacturing of hardware, and the playing of games including hardware energy consumption and Internet traffic.

In our summary table \ref{tab:categories}, we do not include recycling and disposal of hardware and physical copies, nor games-related consumption, practices and communities such as e-sports events, merchandise, and streamers (e.g. on YouTube and Twitch).  But perhaps the biggest omission from our initial calculation in this paper, is mobile gaming.  Gaming on mobile phones represents a large amount of use-time (annual hours played), but with far lower manufacturing impact and use-phase energy.

\begin{table}[t]
    \centering
    \caption{Emissions sources considered in this paper}
    \label{tab:categories}
    \footnotesize
    \setlength{\tabcolsep}{3pt}
    \renewcommand{\arraystretch}{1.1}
    \begin{tabularx}{\columnwidth}{@{}L{0.62}L{1.38}r@{}}
        \toprule
        \textbf{Category} & \textbf{Assumptions and data sources}
            & \textbf{Mt \ceq} \\
        \midrule
        Development and publishing
            & Corporate reports where available (23 companies), omitting Scope~3 use-phase. Estimates for large non-reporting companies (13), based on number of employees or revenue (whichever is lower).
            & \begin{tabular}[t]{@{}l@{\hspace{5pt}}r@{}}
                Reporting & 6.23 \\ Non-reporting & 2.61
              \end{tabular} \\
        \midrule
        Manufacturing
            & Sales for consoles (PS5 \cite{sony_business_data_sales_2026}, Xbox Series~X\textbar S \cite{microsoft_xbox_series_s_2023,microsoft_xbox_series_x_2023,vgchartz_xbox_2026,Lee_2026}, Switch and Switch~2 \cite{nintendo_datasheet,nintendo_IR_historical_data,nintendo_IR_sales_data}). Dell reports \cite{dell2026_reports,Chou_2024,Shilov_2024} for desktops \cite{dell_vostro3030_Carbon_Footprint_2025}, laptops \cite{dell_G15_Carbon_Footprint_2025}, and monitors \cite{dell_u2424he_Carbon_Footprint_2021} are used as category proxies.
            & \begin{tabular}[t]{@{}l@{\hspace{5pt}}r@{}}
                Consoles & 6.21 \\ PCs & 11.52 \\ Monitors & 6.56
              \end{tabular} \\
        \midrule
        Distribution
            & Quantity \cite{sony_business_data_sales_2026,Statista_2025_xbox,nintendo_IR_2026} and download sizes of games \cite{Wikipedia_2026_ps4,Wikipedia_2026_ps5,nintendo_store}. Patches neglected for consoles. Additional carbon to deliver physical copies neglected.
            & \begin{tabular}[t]{@{}l@{\hspace{5pt}}r@{}}
                Steam & 4.38 \\ PS + Xbox & 0.34 \\ Nintendo & 0.01
              \end{tabular} \\
        \midrule
        Online multiplayer
            & Storage, computation, and network costs of serving online multiplayer are neglected. This particularly affects MMOs, shooters/MOBAs, and RTS games.
            & -- \\
        \midrule
        Play time and participation
            & Average power consumption for desktops, laptops, and dominant consoles. Steam Charts for PC playtime  \cite {steam_Charts}. Sony reporting, scaled down for Xbox  \cite{sony2025}. Switch neglected due to reports' lack of differentiation of handheld and docked mode  \cite{nintendo_datasheet}.
            & \begin{tabular}[t]{@{}l@{\hspace{5pt}}r@{}}
                PC & 9.78 \\ Sony ``Game'' & 5.96 \\ Xbox & 1.99
              \end{tabular} \\
        \midrule
        Cloud gaming
            & Industry reports for cloud gaming usage \cite{thakur2026}.  Bitrates of cloud game services~\cite[tab.\,2]{marsden2020}. In-home hardware (display+client) neglected.
            & 6.92 \\
        \midrule
        \textbf{TOTAL} & & \textbf{62.51} \\
        \midrule
        \addlinespace[2pt]
        \multicolumn{3}{@{}l@{}}{\textit{Constants used throughout this paper}} \\
        \multicolumn{3}{@{}l@{}}{Grid carbon intensity: 0.445\,kg~\ceq/kWh (IEA 2024 global average)} \\
        \multicolumn{3}{@{}l@{}}{Download energy intensity: 20\,Wh/GB~\cite{coroama2013-directenergyinternetdataflows}} \\
        \bottomrule
    \end{tabularx}
\end{table}

Our methodology utilises a piecemeal approach drawing on publicly available data from multiple source types to construct emissions estimates across the gaming lifecycle. We draw on corporate sustainability reports and product life cycle assessments (LCAs) published by (\textit{some}) hardware manufacturers, market analyses from industry research firms, publicly available sales and shipment data.  Sources are referenced throughout for reproducibility, and allowing others to revise our estimates as more recent or granular data becomes available.

To derive carbon dioxide equivalent emissions from energy consumption, we utilise the International Energy Agency's 2024 global average grid carbon intensity:  0.445 kg \ceq/kWh.  The IEA forecast is that global average will reduce to about 0.4 kg/kWh by 2027.

\section{Game Development}

Even before a game is distributed or played, there can be significant emissions arising from its development and publishing.  Publishing includes advertising and other steps needed to register the game as a product and market it.  For this component, we use the environmental, social and governance (ESG) reporting from the largest game companies on a 2022 NewZoo list of the largest game companies; to this list we added further reports by snowball collection.  We also include European companies which released reports in response to the European Corporate Sustainability Reporting Directive (CSRD).  The CSRD optionally includes Scope 3 emissions due to the use of products (known as S3.11), and where that is reported we have excluded it these since it should be reflected below in our analysis of annual emissions arising from play (use-time).  We excluded some large companies from our sample, owing to a lack of disaggregated data, or being unrepresentative of typical game developers:  Apple, Aristocrat Leisure, Sony, Nvidia, Microsoft, Google, and Tencent.  Many of these companies do own development studios (typically through acquisitions), but it is difficult to separate game development from their many other activities.

In assembling the data from the ESG and CSRD reporting, we considered annual revenue (converted to US dollars), employees (by headcount), Scope 1 emissions, market-based Scope 2 emissions, and Scope 3 emissions (tracking which categories of Scope 3 are included in the reports).  The companies included, who reported on 2024 GHG emissions are:  Rovio (includes S3.11), Nexon, 37interactive, Paradox Interactive, Take-Two, NCSoft, Nintendo (includes S3.11), Net Ease, Unity, Stillfront (includes S3.11), MTG (includes S3.11), SEGA Sammy, Embracer (includes S3.11), Ubisoft, Nacon, Cybe  rAgent, Electronic Arts, NetMarble, Mixi (includes S3.11), CDProjektRED (includes S3.11), Bandai–Namco, and Capcom (a partial disclosure with no Scope 3 emissions).\footnote{These reports are publicly available on the respective company websites.}

Further, there are significant companies which did \emph{not} disclose in 2024, but which do share employee headcount and revenue totals; using these we can make projections based on industry averages from disclosing companies listed earlier.  We estimated their emissions using the average \ceq intensity on a per employee and a per \$1m USD revenue basis for reporting companies, and applied it to the non-reporting companies, picking the lower of the two.  The companies with incomplete or no GHG disclosures for which we made projections are:  Konami, Nexon, NetMarble, Capcom, Maximum Entertainment, Remedy, Riot Games, Roblox, Square-Enix, Starbreeze, Sybo, Perfect World, and Playtika.  

The reporting companies total 6.934 Mt \ceq, from which we can subtract 0.707 Mt in Scope 3 use-phase emissions for the companies that disclosed these figures in their reports.  For the non-reporting companies, we conservatively compute 2.611 Mt \ceq, choosing the lower estimate for each company out of per-employee and per-unit-revenue.\footnote{If we were to instead choose the higher estimate, the total for the non-reporting companies rises to 4.452 Mt.} This gives a grand total of 8.838 Mt \ceq for game development and publishing by the industry.

\section{Embodied Emissions from Gaming Devices}
\label{sec:hardware}

A primary source of \ceq emissions associated with gaming derives from the manufacture of the hardware required for gameplay conceptualised as the embodied emissions of the device. Dedicated consoles—including the Sony PlayStation, Microsoft Xbox, and Nintendo Switch—are purpose-built for this function, while gaming-configured personal computers constitute a further category. Given the heterogeneity of configurations, we approximate emissions by calculating values for a "proxy" device representative of each category and multiplying by sales volume. In the absence of reliable public data or disclosure regarding sales figures and \ceq for particular hardware categories, we adopt a number of simplifying assumptions to arrive at crude estimates. The resulting figures are intended as indicative approximations and should not be interpreted as fully representative of the categories concerned. Although recent advances in silicon fabrication have extended gaming to conventional laptops and smartphones, we disregard the mobile gaming segment in this calculation. 

For gaming PCs, we disaggregate the segment into desktops and laptops, drawing on Dell's LCA reports and IDC's sales data for 2023 \cite{dell2026_reports,Chou_2024,Shilov_2024}. As no LCA is available for Dell's Alienware range of gaming devices, we use the Dell G15 \cite{dell_G15_Carbon_Footprint_2025} as a laptop proxy since it has similar specifications to the Alienware devices. Manufacturing and transport account for approximately 86.1\% of the G15's total embodied emissions, with reported values ranging from 194 to 766 kg\ceq; we adopt the mean of 308 kg \ceq as a representative figure for an average gaming laptop. For desktops, we use the Dell Vostro 3030 \cite{dell_vostro3030_Carbon_Footprint_2025} as a proxy device based on its specifications, which yields 112 kg \ceq for manufacturing and transport (from a total footprint of 168 kg \ceq). This figure sits at the lower end of the gaming desktop range — high-performance configurations comparable to workstations can exceed 200 kg \ceq — and should therefore be interpreted as a conservative estimate. Because laptops include integrated displays whereas desktops do not, we account separately for monitor emissions using the Dell U2422HE \cite{dell_u2424he_Carbon_Footprint_2021} as a proxy based on its specifications, yielding 405 kg \ceq. IDC reports 44 million gaming PC units sold in 2023, a 13.2\% decline from 2022 and implying approximately 50.7 million units in 2022 \cite{Shilov_2024,Chou_2024}. Applying Market Data Forecast's estimate of a 60\% laptop share yields 30.4 million laptops and 20.3 million desktops \cite{Market_Data_Forecast_2024}. For monitors, back-calculation from IDC's 22.2 million unit forecast for 2024 (a 13.6\% increase) and reported 20.3\% growth in 2023 \cite{Shilov_2024} yields 16.2 million units in 2022. Combined, these volumes give estimated 2022 manufacturing emissions of \textit{9.2 Mt} \ceq for laptops, \textit{2.32 Mt} \ceq for desktops, and \textit{6.56 Mt} \ceq for monitors — a total of approximately \textit{18.08 Mt} \ceq for the gaming PC segment.

For consoles, we apply different strategies across the three main platforms. Sony does not publish manufacturing emissions for its consoles; we assume 200 kg \ceq per unit, a reasonable value given the PS5's weight. Total PS5 sales stand at 92.1 million units, averaging 18.4 million annually over 2020--2025 \cite{sony_business_data_sales_2026}, giving an annual average of 3.68 Mt \ceq. Microsoft reports embodied emissions of 111 kg \ceq for the Xbox Series S and 190 kg \ceq for the Series X \cite{microsoft_xbox_series_s_2023,microsoft_xbox_series_x_2023}. Although Microsoft does not publish sales figures, third-party estimates place total Series sales at around 34 million units \cite{vgchartz_xbox_2026,Lee_2026}, or roughly 6.8 million units annually over 2020--2025. Taking the mean of the two consoles' embodied emissions yields an annual average of approximately 1.02 Mt \ceq. Nintendo similarly does not publish manufacturing emissions, but its CSR data sheet reports total Scope 3 emissions of 3.12 Mt \ceq for fiscal year 2022~\cite{nintendo_datasheet}, during which 17.97 million Switch units were sold~\cite{nintendo_IR_historical_data}. Attributing half of Nintendo's Scope 3 emissions to Switch production gives 87.13 kg \ceq per unit. Cumulative Switch sales of 155.92 million units \cite{nintendo_IR_sales_data} average 17.32 million annually over 2017--2026, yielding an annual average of 1.51 Mt \ceq. Switch 2 sales are aggregated with those of the original Switch.  For simplicity, we treat Switch 2 emissions as equivalent to those of the original Switch.

\section{Distribution}

In this analysis, we neglect the emissions for copies of games sold on physical media, namely Blu-ray (PlayStation and Xbox) and FLASH-based cartridge (Nintendo). We do this for two reasons -- (1) physical distribution of games is becoming marginal, and although any extra impact for distributing games physically will be many more times than distributing digitally, they remain minimal in the scheme of things; (2) available industry sales reports do not distinguish between digitally and physically delivered copies. 

Digital distribution of games results in a data volume transferred, with size depending on the game or update.  A moderate estimate for bulk (large) download energy intensity (including network and data centre, excluding user device) is 20 Wh/GB~\cite{coroama2013-directenergyinternetdataflows}.

\subsection{Steam downloads and updates}

For our estimation, we use an open dataset where bandwidth data was recorded every 10 minutes from SteamDB in GB/s~\cite{han2025steamdownloadbandwidth}. The total data volume was calculated by multiplying the bandwidth value (GB/s) by the duration of the sampling interval (600 seconds [10 minutes]). These volumes were then added together to determine the total Steam data volume:  469,119 PB for 2024.  Multiplying by our constants for download energy intensity and grid intensity yields 4.38 Mt \ceq for Steam.  This includes game downloads, game updates (patches), Steam store browser traffic, and likely multiplayer traffic on Valve games.

\subsection{PlayStation and Xbox}

Sony's sales data indicates that combined PS4 and PS5 software sales ranged from 165.6 million units in FY2020 to 317.9 million units in FY2025 \cite{sony_business_data_sales_2026}, averaging approximately 273.4 million units per year over the six-year period. PS4 game sizes vary widely, from 1 GB to 150 GB \cite{Reddit_2025}, with a simple mean of around 35 GB. However, weighting by sales volume across the list of best-selling titles \cite{Wikipedia_2026_ps4} yields a substantially higher average of 80 GB. PS5 games tend to be larger still: although a comprehensive catalogue of PS5 file sizes is difficult to compile, an unweighted average across the top-selling titles \cite{Wikipedia_2026_ps5} gives approximately 87 GB.  If we assume an average of 87 GB across 273.4 million units, multiplied by our energy intensity for large downloads, we get a total of 0.212 Mt \ceq.

Estimating distribution emissions for Xbox is more difficult, as Microsoft no longer publishes software sales figures. Statista reports 2025 Xbox revenue of \$2.64 billion in retail sales and \$5.91 billion in online sales \cite{Statista_2025_xbox}. Because online revenue includes additional sources such as subscription passes and microtransactions, we assume that half of it derives from game sales. Applying an average price of \$35 per title — reflecting the prevalence of discounted sales — yields an estimated total of approximately 160 million units sold in 2025.  If we assume the download size of an Xbox game is similar to a PS5 game (87 GB), then this equates to 0.124 Mt \ceq.

\subsection{Nintendo}


We took a snapshot of software sales data from Nintendo for the top five Switch 2 games worldwide for April 2025--March 2026.\footnote{The top five as of 31 March 2026 were Mario Kart World, Donkey Kong Bananza, Pokémon Legends: Z-A (Switch 2 Ed.), Pokémon Pokopia and Kirby Air Raiders.} Given that the top five games accounted for more than 56\% of Switch 2 software sales (27.64 million units, against a total of 48.71 million units~\cite[sec.\,1(1)]{nintendo_IR_2026}), we use a simple mean of the game sizes for these games from the Nintendo Store \cite{nintendo_store} to arrive at average game size of 14.08 GB. Multiplying the total sales figure with the average game size, and applying our constants for download energy intensity and grid intensity gives a total of 0.006 Mt, which we round to 0.01 Mt \ceq.

\section{Playing games on PCs and consoles}
\label{sec:playing}
PC gaming happens largely on the Steam platform, and though Steam collects play duration data on the platform it does not share it publicly. Third party websites like Steam Charts~\cite{steam_Charts} aggregate Steam's public API data, however, and a November 2025 scrape~\cite{abraham2026steamstatshoursplayed} provides a basis for an estimation – showing a combined total of 5.57b player hours on Steam for the prior 30 days, across 12k games. Applying a weighted average power consumption figure (305.1 watts) from the Sustainable Games Alliance's Steam hardware model, which combines Steam's Hardware survey with a database of the most popular PC components TDPs, produces a 30-day energy consumption figure of 1.7bn kWh, or 20.7bn kWh if extrapolated to 365 days. Applying a 2024 average global EF results in a figure of 9.779 Mt \ceq.

Sony provides a "Game" within its Scope 3 use-phase emissions, totalling 5.96 Mt \ceq ~\cite{sony2025}. We adopt this figure as representative of the emissions arising from time spent playing on Sony's gaming devices. As Microsoft does not publish comparable data for Xbox, we estimate its use-phase emissions by scaling Sony's figure: given that Xbox Series sales are approximately one-third lower than PS5 sales, we apply a corresponding reduction to obtain an annual impact of 1.987 Mt \ceq.

As mentioned earlier, Nintendo has sold a substantial number of consoles, and aggregate play time is therefore significant. However, the power consumption of the Switch and Switch 2 is considerably lower than that of the PS5 or Xbox Series, particularly when the Switch is used in handheld mode. Refining this estimate would require disaggregated data on hours played in docked versus handheld mode, which is not currently available.

\section{Cloud gaming}

In cloud gaming, games are rendered in a data centre, and the video is streamed to the player.  The graphics processing and related power consumption are thus handled outside the home, and the player minimally needs a display, a relatively thin client (although cloud gaming on PCs and consoles is common), and an input device (usually a game controller).

For this analysis, we neglect the in-home hardware used for playing cloud games.  Either that hardware is a console or PC (with embodied already accounted for in section \ref{sec:hardware}); or it is built into a smart TV or other display; or it is on a small streaming device with relatively low embodied emissions.

This leaves two components for carbon emissions of cloud gaming:  those from energy to render the game on hardware in a data centre; and from the energy required to deliver (stream) the video to the player at home.   To calculate both of these, we need to know how much time people spend playing cloud services.  This is not published by companies offering services (GeForce Now, Xbox Cloud Gaming, PlayStation Plus Premium, and Amazon Luna).  However, industry reports indicate that 400 million or more people use cloud gaming services \cite{thakur2026}.  We might assume these people spend 4.5 hours per week playing cloud games (a median for people that play games on console and PC).  This is an overestimate, since often players' time is split between local PC or console games, and cloud gaming.  But, it is a starting point.

We also need to assign a typical power consumption for cloud rendering, i.e. the consumption of dedicated hardware rendering the game in the data centre.  We choose 150 W, which is on the low side for performance gaming hardware, but we assume that the services use well-optimised hardware and moderate frame rates and resolution.  We hope that data centres where rendering takes place have a significantly lower carbon intensity of electricity supply; this is certainly true where local renewable energy sources (such as geothermal or solar) were provisioned when the data centre was constructed.  For this reason, we use half the world average grid carbon intensity from table~\ref{tab:categories} for cloud rendering.  A further improvement would be to apply carbon intensities by region, depending on the carbon intensity of their data centre electricity supply.  However, hours played and cloud subscriptions are not available by region, and companies do not publish the locations of data centres where cloud rendering take place.

To estimate the delivery component, we use the Carbon Trust 2021 estimate of 0.027 kg CO2e per hour of streaming \cite{carbontrust2021}.  However, we increase this by 50\% to 0.041 kg CO2e/h, since according to prior measurements~\cite[tab.\,2]{marsden2020} cloud gaming has a higher bandwidth usage (10--40 Mbps) than watching video streams (2--16 Mbps).

These result in a yearly 3.124 Mt \ceq for rendering, and 3.791 Mt for delivery --- a total of 6.915 Mt for cloud gaming as a whole.

\section{Discussion}
\label{sec:discussion}

Our intent with this short paper is to spark further discussion on sustainability in and of the gaming industry. Even a crude annual estimate of the sector's emissions surfaces important implications and invites scrutiny of data and methods. Chief among these is the need for more transparent and standardised reporting from industry actors. As our crude estimates make clear, considerable uncertainty stems from the patchy availability of disaggregated data — particularly around gaming-specific hardware, regional manufacturing emissions, and use-phase electricity consumption. Improved disclosure would enable more precise accounting and support meaningful comparison across platforms, manufacturers, and scenarios.

A related point is that hardware and embodied emissions may already be receiving some attention, both within the gaming industry and the broader consumer electronics sector where lifecycle reporting is increasingly standardised. However, this picture is uneven. While actors such as Dell and Apple regularly publish LCAs and environmental reports, many other manufacturers do not. The emissions associated with console manufacture are conspicuously absent from the public record, despite consoles being produced by large firms operating at a scale comparable to Dell. Mid-generation console refreshes effectively introduce additional manufacturing cycles within a single console generation, challenging the assumption that embodied emissions are well-bounded. A similar dynamic applies to gaming PCs, where frequent component upgrades, particularly GPUs, propel embodied emissions that fall outside the typical "one device per user per generation" accounting frame. We acknowledge the crude nature of our estimates and the debatable status of some of our assumptions, and we treat this as an open invitation to the community to refine the methodology.

Continued and arguably more urgent attention to the use phase of gaming devices also appears warranted and this emphasis is increasingly reflected in industry practice as more actors are now forthcoming with the energy consumption of various devices. The use phase remains the largest source of gaming-related emissions for most platforms, and its trajectory is shaped by ongoing developments that existing scenarios may not fully capture. Chief among these is the trajectory of cloud gaming. Its emissions profile depends heavily on the endpoint device: thin clients in the home can reduce overall embodied emissions and home energy, but if cloud gaming is played on full-spec consoles, PCs, or gaming laptops, the net effect could be additive rather than substitutive. The increasing integration of AI — not only in development pipelines but within games themselves (for rendering, NPC behaviour, procedural content generation) — represents another use-phase variable whose energy implications are not yet well-characterised.

Our global annual total (62.5 Mt \ceq) is an order of magnitude lower than the Marsden et al.\,\cite{marsden2020} 2030 forecast for any of the three scenarios (432--570 Mt \ceq).  This is mostly because our energy and emissions factors for data traffic (downloading and cloud streaming) are much lower, based on publications since 2020.  Moreover, we can think further on the scenarios developed by Marsden et al. \cite{marsden2020}. It is worth asking whether current developments call for an additional fourth scenario, or whether what we are observing is better described as a hybrid of traditional, locally rendered games, with some uptake of cloud gaming (their Scenario 3). Either way, the underlying scenarios merit re-examination in light of cloud gaming uptake, AI integration, and hardware upgrade cycles that the original model may not adequately reflect.

\textbf{Acknowledgements.}  There has been dedicated master thesis project work by Basharat Mahmood~\cite{mahmood2025}, Rahul Tikkavarapu~\cite{tikkavarapu2026} and Jonas Tysk Hedlund~\cite{hedlund2026}.  While their analyses were necessarily more partial and they were not involved as authors here, our discussions with them about factors to account for and appropriate accounting methods, contributed to our development of this paper.

\bibliographystyle{ACM-Reference-Format}
\bibliography{_gamesLCA}

\end{document}